\documentclass{ifacconf}
\usepackage{amsmath}
\usepackage{amstext}
\usepackage{amssymb}
\usepackage{tikz}
\usepackage{color}
\usepackage{amsfonts}
\usepackage{amsmath}
\usepackage{arydshln}
\usepackage{tikz}
\usepackage{overpic}
\usepackage{multicol}
\usepackage{enumitem}
\usepackage{graphicx}      
\usepackage{natbib}        

\usepackage{xparse}
\usepackage{mathdots}
\DeclareMathAlphabet\mathbfcal{OMS}{cmsy}{b}{n}

\begin{document}
\begin{frontmatter}

\title{A Networked Competitive Multi-Virus SIR  Model: Analysis and Observability}

\thanks[footnoteinfo]{Research supported in part by the National Science Foundation, grants NSF-ECCS \#2032258 and NSF-ECCS \#2032321.}

\author[First]{Ciyuan Zhang} 
\author[Second]{Sebin Gracy} 
\author[Third]{Tamer Ba\c sar}
\author[First]{Philip E. Par\'e}

\address[First]{
Elmore Family School of Electrical and Computer Engineering at Purdue University (email: \{zhan3375,philpare\}@purdue.edu)}
\address[Second]{
School of Electrical Engineering and Computer Science, KTH Royal Institute of Technology, Stockholm, Sweden (email: gracy@kth.se)}
\address[Third]{Coordinated Science Laboratory, University of Illinois at Urbana-Champaign (email: basar1@illinois.edu)}

\begin{abstract}                
This paper proposes a novel discrete-time multi-virus SIR (susceptible-infected-recovered) model that captures the spread of competing SIR epidemics over a population network. First, we provide a sufficient condition for the infection level of all the viruses over the networked model to converge to zero in exponential time. Second, we propose an observation model which captures the summation of all the viruses' infection levels in each node, which represents the individuals who are infected by different viruses but share similar symptoms. We present a sufficient condition for the model to be locally observable. We propose a Luenberger observer for the system state estimation and show via simulations that the estimation error of the Luenberger observer converges to zero before the viruses die out. 
\end{abstract}

\begin{keyword}
Biological networks and epidemics dynamics
\end{keyword}

\end{frontmatter}

\vspace{-1ex}
\section{Introduction}

\vspace{-2ex}

The history of human civilization has been a narrative of undergoing, battling, and outmatching various pandemics~\citep{benedictow2004black, johnson2002updating}. 
Suffering severe life and economic loss, the research of modeling and monitoring the spread of multiple diseases concurrently has grown significantly through the inspection of each epidemic. In this paper, we investigate the modeling, analysis, and observation of the spread of multi-viruses over population networks.

A considerable amount of effort has been expended on the study of
multi-virus models~\citep{pare2017multi, prakash2012winner, sahneh2014competitive, pare2020analysis, santos2015bi, liu2016analysis, pare2021multi}, which focus on the competing susceptible-infected-susceptible (SIS) networked virus model. 
In this paper our focus is on the competing susceptible-infected-recovered (SIR) epidemic model over a network, as the SIR model can capture
the behavior of a diverse set of different epidemics
such as: H1N1~\citep{coburn2009modeling}, Ebola~\citep{berge2017simple}, and COVID-19~\citep{chen2020time}. 
The single virus SIR epidemic networked model has been studied extensively, e.g.,~\citep{hota2021closed, mei2017dynamics, pare2020modeling}. However, to the best of our knowledge, the competing SIR epidemics has not been studied in the literature. Thus, in this work we propose a discrete-time competing SIR virus networked model. The multi-virus model captures the presence and spread of viruses such as influenza and the SARS-CoV-2 virus over a population and could also be utilized to represent different behaviors of variants of the SARS-CoV-2 virus~\citep{lopez2021effectiveness}.


Beyond the modeling and analysis of the epidemic models, the epidemic monitoring and infection level estimation have been crucial to the research on contagions. Given that the SARS-CoV-2 pandemic has provided us with an enormous amount of data, how to accurately infer the infection levels of the infectious diseases has become a topic requiring urgent attention~\citep{barmparis2020estimating, meyerowitz2020systematic}. However, the various symptoms caused by diseases such as influenza~\citep{monto2000clinical} and SARS-CoV-2~\citep{tostmann2020strong} affect the measurement of the cases of different diseases and pose difficulties for the estimation of the states of different epidemics, especially when tests are limited as was witnessed at the beginning of the pandemic and at various peaks of different waves. 

In this paper, we propose what we believe to be the first multi-virus model of SIR networked epidemic spreading, along with specifications that ensure the model is well defined. 
We then provide sufficient conditions for the infection level of each virus to converge to zero in exponential time. 
Moreover, we explore the system state estimation with an observation model which captures the summation of all cases that exhibit similar signs of illness.



\vspace{-1ex}
\subsection{Notation}
\vspace{-2ex}

We denote the set of real numbers and the set of non-negative integers by $\mathbb{R}$ and $\mathbb{Z}_{\geq 0}$, respectively. For any positive integer $n$, we have $[n]:=\{1,2,...,n \}$. The spectral radius and an eigenvalue of a matrix $A \in \mathbb{R}^{n\times n}$ are denoted by $ \rho(A)$ and $\lambda(A)$, respectively. A diagonal matrix is denoted by diag$(\cdot)$. The transpose of a vector $x\in \mathbb{R}^n$ is $x^\top$. The Euclidean norm is denoted by $\lVert \cdot \rVert$. We use $I$ to denote the identity matrix. We use $\mathbf{0}$ to denote the vectors whose entries are all 0, where the dimensions of the vectors are determined by context. Given a matrix $A$, $A \succ 0$ 
indicates that $A$ is positive definite, 
whereas $A \prec 0$ 
indicates that A is negative definite.


 




\vspace{-1ex}

\section{Background}\label{sec:background}
\vspace{-2ex}
In this section, we present our system model, the set of questions to be addressed, and some auxiliary results to be used in subsequent sections.


\vspace{-1ex}
\subsection{System Model}
\vspace{-2ex}

\begin{figure}
\centering
\includegraphics[width=.3\textwidth]{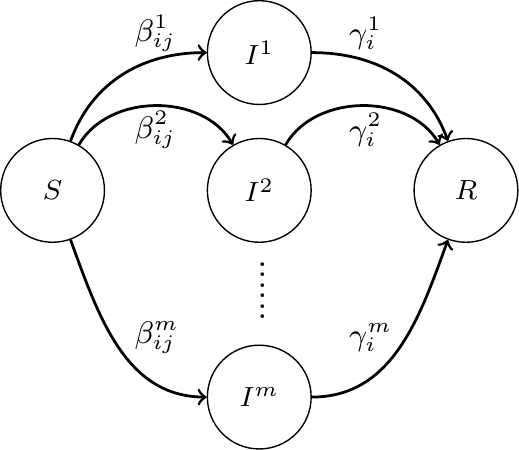}
\caption{The competing SIR networked model. Each node in the network can only be in one of three states: $S$, $I^k$, $R$, where $k \in[m]$.}
\label{fig:SI^nR}
\end{figure}

We consider a discrete-time dynamics for the networked model of the multi-virus SIR epidemics. There are $m$ viruses spreading over the network and each individual can be infected by no more than one virus. We denote by $\beta^k_{ij}$ the infection rate of the $k$-th virus from node $j$ to node $i$, and by $\gamma^k_i$ the healing rate for node $i$ with respect to virus $k$. We denote by $s_i$ and $r_i$ the susceptible and recovered proportions of subpopulation $i$, respectively. We use $x^k_i[t]$, where $k \in [m]$, to denote the fraction of individuals infected with virus $k$ at time instant $t$.  A graphical depiction of this
model is given in Figure~\ref{fig:SI^nR}. The discrete-time dynamics of the time-invariant competing virus of SIR networked epidemic model are written as:
\vspace{-0.5ex}
\begin{subequations}
\label{eq:dt_SIR} 
\begin{align}
  &s_i[t+1] = s_i[t] -h\Bigg(s_i[t]\sum_{k=1}^m\sum_{j=1}^n \beta^k_{ij} x^k_j[t]\Bigg), \label{eq:dt_SIRsub1}\\
 &x^k_i[t+1] = x^k_i[t]+ h\Bigg(s_i[t]\sum_{j=1}^n \beta^k_{ij} x^k_j[t]-\gamma^k_i x^k_i[t]\Bigg),  \label{eq:dt_SIRsub2}
\\
 &r_i[t+1] = r_i[t] +h\sum_{k=1}^m \gamma^k_i x^k_i[t], \label{eq:dt_SIRsub3}
\end{align}
\end{subequations}
\vspace{-2.75ex}

\noindent
where $h>0$ is the sampling parameter, 
$t$ is the time index, and $k\in [m]$ indicates the $k$-th virus. Notice that $s_i[t] +x_i^1[t] + \cdots+x_i^m[t]+r_i[t]= 1$, capturing the fact that in the competing virus scenario, all the viruses are exclusive: an individual cannot be infected by more than one virus concurrently. We now rewrite~\eqref{eq:dt_SIRsub2} as:
\begin{equation}\label{eq:SIR_dynamic}
    x^k[t+1] =x^k[t] +h \{S[t]B^k - \Gamma^k\}x^k[t],
\end{equation}
where $S[t] = $ diag$(s_i[t])$, 
$B^k$ is a matrix with ($i,j$)-th entry $\beta^k_{ij}$, and $\Gamma^k[t] =$ diag$(\gamma^k_i)$.

We now introduce the following assumptions to ensure that the model in~\eqref{eq:dt_SIR} is well-defined. 

\begin{assum}\label{assume:one}
For all $i \in [n]$ and $k \in [m]$, we have $s_i[0], x^k_i[0], r_i[0] \in [0,1]$.
\end{assum}

\begin{assum}\label{assume:two}
For all $i \in [n]$, and $k \in [m]$, we have $\beta^k_{ij} \geq 0, \gamma^k_i> 0$.
\end{assum}

\begin{assum}\label{assume:three}
For all $i \in [n]$,  and $k \in [m]$, we have $h\sum_{k=1}^m \sum_{j=1}^n \beta^k_{ij} \leq 1$ and $ h\sum_{k=1}^m \gamma^k_i\leq 1$.
\end{assum}

\begin{rem}
Assumptions~\ref{assume:one} and \ref{assume:two} can be interpreted as the initial proportion of susceptible, infected, and recovered individuals all 
lying
in the interval of $[0,1]$ and we assume that the healing rates are always positive, which are both reasonable. Assumption~\ref{assume:three} ensures the sampling rate is frequent enough for the states of the model to remain well defined.
\end{rem}



We next build an observation model which produces the output as the proportion of individuals who show flu-like symptoms from infection of all viruses. The observation model is written as  (where we repeat \eqref{eq:dt_SIRsub2} for convenience):
\vspace{-3ex}
\begin{subequations}
\label{eq:SIR_observation} 
\begin{align}
  x^k_i[t+1] &= x^k_i[t] +h\bigg\{s_i[t]\sum_{j=1}^n \beta^k_{ij} x^k_j[t]-\gamma^k_i x^k_i[t]\bigg\}, \label{eq:SIR_observationsub1}\\
y_i[t] & = \sum_{k=1}^m c_i^k x_i^k[t],\label{eq:SIR_observationsub2}
\end{align}
\end{subequations}
where $c_i^k$ is the measurement coefficient.
\begin{assum}\label{assume:four}
The coefficient $c_i^k\in (0,1]$ for all $i\in [n], k\in[m]$.
\end{assum}
\begin{rem}
The coefficient $c_i^k$ 
from Eq.~\eqref{eq:SIR_observationsub2} can capture the probability of showing symptoms from the $k$-th virus at subpopulation~$i$. Therefore, $1-c_i^k$ captures the probability of individuals infected with the $k$-th virus in subpopulation $i$
being asymptomatic.
The coefficient $c_i^k $ can also represent how each subpopulation $i$ defines and measures the cases based on the symptoms of each virus $k$. For example, the symptoms of influenza can include but are not limited to fever, muscle aches, cough, runny nose, headaches, fatigue, etc.
\end{rem}

\begin{rem}
In the observation model, Eq.~\eqref{eq:SIR_observationsub2} can be interpreted as the summation of all the number of symptomatic patients in each subpopulation, which is an indicator for the decision-makers to be able to judge adequacy of the local hospital capacity and the availability of medical resources against the need. 
\end{rem}
\noindent We then have the following results for the system model under Assumptions~\ref{assume:one}-\ref{assume:three}. 

\begin{lem}\label{lemma:one}
Suppose that $s_i[0], x^k_i[0], r_i[0] \in [0,1]$, $s_i[0]+\sum_{k=1}^m x^k_i[0]+r_i[0] =1$ for all $k\in[m], i\in[n]$, and Assumptions~\ref{assume:one}, \ref{assume:two}, and \ref{assume:three} hold. Then, for all $i \in[n]$ and $t \in \mathbb{Z}_{\geq0}$,
\begin{enumerate}
    \item $s_i[t], x^k_i[t], r_i[t] \in [0,1]$, for all $k \in [m]$, and $s_i[t]+\sum_{k=1}^m x^{k}_i[t]+r_i[t] =1$, and 
    \item $s_i[t+1] \leq s_i[t]$.
\end{enumerate}
\end{lem}
\vspace{-2ex}
\textit{Proof:}
  1) We prove this result by induction. 
\noindent \textit{Base Case:}
By the assumptions made, $s_i[0], x^k_i[0], r_i[0] \in [0,1]$, $s_i[0]+\sum_{k=1}^m x^k_i[0]+r_i[0] =1$ for all $k\in[m], i\in[n]$. From Assumptions~\ref{assume:one}-\ref{assume:three}, we know that $s_i[0]\geq 0$ and $1-h\sum_k^m \sum_j^n \beta_{ij}^k x_j^k[0] \geq0$, hence $s_i[1] = s_i[0](1-h\sum_k^m \sum_j^n \beta_{ij}^k x_j^k[0])\geq0$. Since $-h(s_i[t]\sum_k^m \sum_j^n \beta_{ij}^kx_j[0])\leq 0$, we obtain that $s_i[1]\leq s_i[0]\leq 1$. 
We can also acquire that $x_i^k[1]\geq x_i^k[0](1-h\gamma_i^k)\geq 0$ and $x_i^k[1]\leq x_i^k[0]+hs_i[0]\sum_j^n \beta_{ij}^k x_j^k[0]\leq x_i^k[0]+s_i[0]\leq 1$. Ultimately, we have $r_i[1]\geq r_i[0]\geq0$ and $r_i[1]\leq r_i[0]+ \sum_k^m x_i^k[0]\leq 1$. Summing up Eqs.~\eqref{eq:dt_SIRsub1}-\eqref{eq:dt_SIRsub3}, we obtain that $s_i[1]+\sum_{k=1}^m x^{k}_i[1]+r_i[1] =s_i[0]+\sum_{k=1}^m x^{k}_i[0]+r_i[0] =1$.

\noindent \textit{Inductive Step:} We assume for some arbitrary $t$ that the following holds: $s_i[t], x^k_i[t], r_i[t] \in [0,1]$, for all $k \in [m]$ and $s_i[t]+\sum_{k=1}^m x^{k}_i[t]+r_i[t] =1$. By repeating the same steps from the \textit{Base Case} except replacing $0$ and $1$ with $k$ and $k+1$, we can write that $s_i[t+1], x^k_i[t+1], r_i[t+1] \in [0,1]$, for all $k \in [m]$ and $s_i[t+1]+\sum_{k=1}^m x^{k}_i[t+1]+r_i[t+1] =1$. Therefore, by induction, we can prove that $s_i[t], x^k_i[t], r_i[t] \in [0,1]$, for all $k \in [m]$ and $s_i[t]+\sum_{k=1}^m x^{k}_i[t]+r_i[t] =1$ for all $i \in[n]$ and $t\in  \mathbb{Z}_{\geq0}$.

2) From 1) and Assumption~\ref{assume:two} we know that \\  $-h\Big(s_i[t]\sum_{k=1}^m\sum_{j=1}^n \beta^k_{ij} x^k_j[t]\Big)\leq 0$ for all $t\in  \mathbb{Z}_{\geq0}$. Thus, we have $s_i[t+1]\leq s_t[t]$ for all $i\in [n]$ and $t\in  \mathbb{Z}_{\geq0}$. 

\vspace{-1ex}
\subsection{Problem Formulation}

\vspace{-2ex}

With the model in place, we now introduce the problems considered in this paper under Assumptions~\ref{assume:one}-~\ref{assume:four}:


\begin{enumerate}[label=(\roman*)]
    \item \label{itm:first} For the system with dynamics given in~\eqref{eq:SIR_dynamic}, provide a sufficient condition which ensures that $x^k[t]$ for some and all $k\in[m]$ converges to the eradicated state, namely $x^k=\mathbf{0}$, in exponential time.
    \item What is the rate of convergence for the sequence $x^k[t], k\in[m]$ (converging to $\mathbf{0}$ exponentially)?
    \item \label{itm:third} 
    Given the observation $y_i[t]$, 
    under what conditions are the infection levels of each virus $x_i^k[t]$, for all $i\in [n], k \in [m] $ locally observable, at $s_i[t]=0, \forall i \in [n]$?
\end{enumerate}

\vspace{-1ex}
\subsection{Preliminaries} 
\vspace{-2ex}

Consider a system described as follows:
\begin{subequations}
\label{eq:dt_prelim} 
\begin{align}
  x[t+1] &= f(t,x[t]), \label{eq:dt_prelimsub1}\\
 y[t] & = g(x[t]). \label{eq:dt_prelimsub2}
\end{align}
\end{subequations}

\begin{defn}\label{def:GES}
An equilibrium point of \eqref{eq:dt_prelimsub1} is GES if there exist positive constants $\alpha$ and $\omega$, with $0\leq \omega <1$, such that
\begin{equation}
    \lVert x[t] \rVert \leq \alpha \lVert x[t_0] \rVert \omega^{(t-t_0)}, \forall t\geq t_0 \geq 0, \forall x[t_0] \in \mathbb{R}^n.
\end{equation}
\end{defn}
\begin{lem}\label{lemma:GES}
\cite[Theorem 28]{vidyasagar2002nonlinear}
Suppose that there exist a function $V: \mathbb{Z}_+ \times \mathbb{R}^n \rightarrow \mathbb{R}$, and constants $a,b,c>0$ and $p>1$ such that $a\lVert x \rVert^p \leq V(t,x) \leq b \lVert x \rVert^p$, $\Delta V(t,x):= V(x[t+1])-V(x[t]) \leq -c\lVert x \rVert^p, \forall t \in \mathbb{Z}_{\geq0}$. Then $\forall x (t_0)\in \mathbb{R}^n$, $x=\mathbf{0}$ is the globally exponential stable equilibrium of \eqref{eq:dt_prelimsub1}. 
\end{lem}
\begin{lem}\label{lemma:rate_GES}
\cite[Theorem 23.3]{rugh1996linear} Under the conditions of Lemma~\ref{lemma:GES}, convergence to 
the origin 
has an exponential rate of at least $\sqrt{1-(c/b)} \in [0,1)$, where $b$ and $c$ are as defined in Lemma~\ref{lemma:GES}.
\end{lem}
\begin{lem}\label{lemma:diagonal}
\cite[Proposition 2]{rantzer2011distributed}
Suppose that $M$ is a nonnegative matrix which satisfies $\rho(M)<1$. Then there exists a diagonal matrix $P \succ 0$ such that $M^\top P M -P \prec 0$. 
\end{lem}
 
\begin{defn}
The system in Eq.~\eqref{eq:SIR_observation} is locally observable at 
$s[t]$
if we are able to recover 
$x_i^k [t]$ for all $i \in [n], k\in [m]$
through the output in the duration of $[t,t+m-1]$.
\end{defn}
 
\begin{lem}\citep{sontag1979observability}\label{lemma:locally_observable}
The system in~\eqref{eq:dt_prelim} is locally 
observable at $x[t]$ 
if and only if the map $x[t] \rightarrow (g(x[t]),g(f^1(x[t])),\cdots, g(f^{n-1}(x[t])))$ 
is injective, where $n$ is the dimension of $x[t]$.
\end{lem}



 

\vspace{-1ex}
\section{Healthy State Analysis} \label{sec:stability}

\vspace{-2ex}

This section presents sufficient conditions that guarantee
that each virus $k$ converges to zero exponentially fast, and provides the associated rates of convergence for each virus. We then present the conditions for all the viruses to converge to zero in exponential time. 
Similar to the 
standard
SIR 
model, the multi-competitive SIR networked model converges to a healthy state regardless of the system parameters and initial conditions; however, it is important to study the exponential convergence as it guarantees that the viruses die out at a faster rate and fewer individuals become infected over the course of the outbreak.

Let
\vspace{-2ex}
\begin{align}
     M^k &:= I-h\Gamma^k +hB^k, \label{eq:M} \\
     \Tilde{M}^k[t]  &:= I+ h\{S[t]B^k - \Gamma^k\},
    \label{eq:Mhat}
\end{align}
and note that $\Tilde{M}^k $ is the state transition matrix of Eq.~\eqref{eq:SIR_dynamic}:
\begin{equation}\label{eq:M_M_hat}
    \Tilde{M}^k[t] = M^k-h(I-S[t])B^k.
\end{equation}
We first present a sufficient condition, in terms of $M^k$, for the viruses to converge to zero exponentially.


\begin{thm}\label{thm:GES}
Under Assumptions \ref{assume:one}-\ref{assume:three}, if $\rho(M^k)< 1$, 
then the $k$-th virus of the system in \eqref{eq:dt_SIR} converges to zero in exponential time, and this holds for all $k\in [m]$.
\end{thm}
\vspace{-2.5ex}
\textit{Proof:}
Consider the candidate Lyapunov function: $V^k(t,x^k) = (x^k)^{\top} P^kx^k$. Since $P^k$ is diagonal and positive definite, $(x^k)^\top P^k x^k>0$, for all $x^k \neq \mathbf{0}$. 
Therefore, $V^k(t,x^k)>0$ for all $k\in[m], t \in \mathbb{Z}_{\geq 0}$, $x^k \neq \mathbf{0}$. 
Since $P^k$ is positive definite, 
\begin{equation}
  \lambda_{\text{min}}(P^k)I \leq P^k \leq  \lambda_{\text{max}}(P^k)I,    
\end{equation}
which implies that
\begin{equation}\label{eq:sigma2}
    \sigma_1^k\| x^k\|^2 \leq V^k(t,x^k) \leq  \sigma_2^k \| x^k\|^2,
\end{equation}
where 
$\sigma_1^k = \lambda_{\text{min}}(P^k)$ and $\sigma_2^k=  \lambda_{\text{max}}(P^k)$, with $\sigma_1^k, \sigma_2^k>0$.

Now we turn to computing $\Delta V^k(t,x^k)$. For $x^k\neq 0$ and for all $k \in[m]$, using \eqref{eq:SIR_dynamic} and \eqref{eq:M}-\eqref{eq:Mhat}, we have
\begin{align}
    &\Delta V^k(t,x^k) \nonumber \\ &
    = 
    (x^k)^\top \Tilde{M}^k [t]^\top P^k\Tilde{M}^k[t]x^k- (x^k)^\top P^k x^k \nonumber \\ 
    &= 
    (x^k)^\top [(M^k)^\top  P^kM^k- P^k]x^k \nonumber \\ & \ \  \ \ -2h (x^k)^\top (B^k)^\top (I-S[t])P^k M^k x^k \nonumber \\ 
    & \ \  \ \  +h^2 (x^k)^\top (B^k)^\top (I-S[t])P^k(I-S[t])B^k x^k.\label{eq:deltaV}
\end{align}
Note that the second and third terms of \eqref{eq:deltaV} can be reorganized as 
\begin{align}
    &(x^k)^\top [-2h (B^k)^\top (I-S[t])P^k M^k \nonumber \\&\ \   \ \   \ \ 
    +h^2 (B^k)^\top (I-S[t])P^k(I-S[t])B^k ] x^k \nonumber \\
    &= (x^k)^\top \{h (B^k)^\top (I-S[t])P^k\nonumber \\
    & \ \  \ \ [-2M^k+ h(I-S[t]) B^k]\}x^k\nonumber \\
    &= (x^k)^\top \{h (B^k)^\top (I-S[t])P^k\nonumber \\ \label{eq:second_third_term}
    & \ \  \ \ [-2(I -h\Gamma^k[t])-h(I+S[t])B^k ]\}x^k\leq 0, 
\end{align}
where the last equality follows from \eqref{eq:M}, and the inequality follows from Assumptions \ref{assume:two}-\ref{assume:three} and Lemma~\ref{lemma:one}. 
Thus, by plugging~\eqref{eq:second_third_term} into~\eqref{eq:deltaV}, we obtain
\begin{equation}\label{eq:lyap_thirdterm}
    \Delta V(t,x^k) \leq (x^k)^\top [(M^k)^\top  P^k M^k- P^k]x^k.
\end{equation}
Since $[(M^k)^\top  P^kM^k- P^k]$ is negative definite, 
we have, from Eq.~\eqref{eq:lyap_thirdterm},
\begin{equation}\label{eq:sigma3}
    \Delta V(t,x^k) \leq - \sigma_3^k \| x^k\|^2,
\end{equation}
where $\sigma_3^k =  \lambda_{\text{min}}[P^k-(M^k)^\top  P^kM^k]$, with $\sigma_3^k>0$.

Therefore, from~\eqref{eq:sigma2} and \eqref{eq:sigma3}, $V^k(t,x^k)$ is a Lyapunov function, with an exponential decay, and hence, $x^k$ converges to zero at an exponential rate.
%
%



\begin{cor}\label{coro:rate_exp}
Under the assumptions of Theorem~\ref{thm:GES}, and with $P^k$ as defined in the proof of Theorem 1, convergence of $x^k[t]$ generated by Eq.~\eqref{eq:SIR_dynamic} has an exponential rate of at least $\sqrt{1-\frac{\sigma_3^k}{\sigma_2^k}}$, where $\sigma_2^k= \lambda_{\text{max}}(P^k)$, $\sigma_3^k = \lambda_{\text{min}}[P^k-(M^k)^\top  P^k M^k]$ for each $k \in[m]$.
\end{cor}
\textit{Proof:}
From Lemma~\ref{lemma:rate_GES}, \eqref{eq:sigma2}, and \eqref{eq:sigma3}, the rate of convergence of virus $k$ is upper bounded by $\sqrt{1-\frac{\sigma_3^k}{\sigma_2^k}}$.  We then need to show that the rate is well defined, which is $\sqrt{1-\frac{\sigma_3^k}{\sigma_2^k}}  \in [0,1)$. Since $\sigma_2^k>0$ and $\sigma_3^k>0$, it will be sufficient to  show that $\sigma_2^k \geq \sigma_3^k$.

Since $P^k$ is positive definite and $(M^k)^\top P^k M^k $ is nonnegative definite, we have
\begin{equation}
    \sigma_3^k I \leq P^k - (M^k)^\top P^k M^k \leq P^k \leq \sigma_2^k I,
\end{equation}
from which $\sigma_2^k\geq \sigma_3^k$ and the rate of convergence $\sqrt{1-\frac{\sigma_3^k}{\sigma_2^k}}$ is well defined.
%

\vspace{-2ex}

\section{State Observation Model}\label{sec:observation}

\vspace{-2ex}

In this section, we use the measurement of $y_i[t]$, the fraction of individuals who show similar symptoms from all viruses, to determine the infection level of each virus. We first construct the observability matrix for the system from Eq.~\eqref{eq:SIR_observationsub2}, 
writing Eq.~\eqref{eq:SIR_observationsub2} as:
\begin{equation}\label{eq:observer_y_full}
    \mathbf{y}[t] = \mathbf{C} \mathbf{X}[t],
\end{equation}
where $\mathbf{y}[t] = \begin{bmatrix}
y_1[t] &
y_2[t] &
\cdots &
y_n[t]
\end{bmatrix}^\top \in \mathbb{R}^{n\times 1}$, the measurement matrix $\mathbf{C} \in \mathbb{R}^{n\times mn}$ is:

\begin{equation}
    \mathbf{C} = \begin{bmatrix}
C^1  & C^2& \cdots & C^m  
\end{bmatrix} \nonumber 
\end{equation}
with $C^k = \text{diag}([c_1^k, c_2^k, \cdots, c_n^k])$ for all $k \in [m]$,
$\mathbf{X}[t] \in \mathbb{R}^{mn\times 1}$ is:
\begin{equation}
    \mathbf{X}[t] = \begin{bmatrix}
x^1[t] \\
x^2 [t]\\
\vdots \\
x^m[t]
\end{bmatrix}. \nonumber 
\end{equation}
Therefore, the measurement $\mathbf{y}[t] $ can be reorganized as:
\begin{align}
     \mathbf{y}[t]  
= C^1x^1[t]+C^2x^2[t]+\cdots+ C^mx^m[t].
\end{align}
We can express the measurement at each time step 
over the time interval $[t,t+m-1]$ as:
\vspace{-1ex}
\begin{align}
    & \begin{bmatrix}
\mathbf{y}[t] \\
\mathbf{y}[t+1] \\
\mathbf{y}[t+2]\\
\vdots \\
\mathbf{y}[t+m-1]
\end{bmatrix}\nonumber \\ = &
\begin{bmatrix}
C^1x^1[t]+\cdots+ C^mx^m[t] \\
C^1x^1[t+1]+\cdots+ C^mx^m[t+1]  \\
C^1x^1[t+2]+\cdots+ C^mx^m[t+2] \\
\vdots \\
C^1x^1[t+m-1]+\cdots+ C^mx^m[t+m-1] 
\end{bmatrix}\nonumber 
\\ 
=& 
\begin{bmatrix}
C^1 \\
C^1\Tilde{M}^1[t]  \\
C^1\Tilde{M}^1[t]\Tilde{M}^1[t+1] \\
\vdots \\
C^1\Tilde{M}^1[t]\cdots \Tilde{M}^1[t+m-2]
\end{bmatrix}x^1[t] \nonumber  \\&+ \begin{bmatrix}
C^2  \\
C^2\Tilde{M}^2[t]  \\
C^2\Tilde{M}^2[t]\Tilde{M}^2[t+1] \\
\vdots \\
C^2\Tilde{M}^2[t]\cdots \Tilde{M}^2[t+m-2]
\end{bmatrix}x^2[t]+ \cdots \nonumber 
\\
&  + \begin{bmatrix}
C^m  \\
C^m\Tilde{M}^m[t]  \\
C^m\Tilde{M}^m[t]\Tilde{M}^m[t+1] \\
\vdots \\
C^m\Tilde{M}^m[t]\cdots \Tilde{M}^m[t+m-2]
\end{bmatrix}x^m[t]\nonumber  
\\  =& \mathcal{O}^1[t] x^1[t]+ \mathcal{O}^2[t] x^2[t]+ \cdots + \mathcal{O}^m[t] x^m[t]\nonumber \\
= &
\begin{bmatrix}
\mathcal{O}^1[t] &
\mathcal{O}^2[t] &
\cdots &
\mathcal{O}^m[t]
\end{bmatrix}\mathbf{X}[t]
,\label{eq:Ax=b}
\end{align}
where
\begin{equation}
    \mathcal{O}^k[t] =  \begin{bmatrix}
C^k  \\
C^k\Tilde{M}^k[t]  \\
C^k\Tilde{M}^k[t]\Tilde{M}^k[t+1] \\
\vdots \\
C^k\Tilde{M}^m[t]\cdots \Tilde{M}^k[t+m-2]
\end{bmatrix}, \nonumber 
\end{equation}
with $\mathcal{O}^k[t]  \in \mathbb{R}^{mn\times n}$ for all $k \in [m]$.
We now define the observability matrix of the system in Eq.~\eqref{eq:SIR_observation} as:
\begin{equation}
    \mathbb{O}[t] = \begin{bmatrix}
\mathcal{O}^1[t] &
\mathcal{O}^2[t] &
\cdots &
\mathcal{O}^m[t]
\end{bmatrix}, \label{eq:observability_matrix}
\end{equation}
where $\mathbb{O}[t] \in \mathbb{R}^{mn\times mn}$.



We now consider the case when $s_i[t] =0, \forall i \in [n]$. 
Then the observability matrix in Eq.~\eqref{eq:observability_matrix} becomes 
\begin{equation}
    \mathbb{O}_0[t] = \begin{bmatrix}
\mathcal{O}_0^1[t] &
\mathcal{O}_0^2[t] &
\cdots &
\mathcal{O}_0^m[t]
\end{bmatrix}, \label{eq:observability_matrix_zero}
\end{equation}
where
\vspace{-1ex}
\begin{equation}
    \mathcal{O}_0^k[t] =  \begin{bmatrix}
C^k  \\
C^k(I-h\Gamma^k)  \\
C^k(I-h\Gamma^k)^2 \\
\vdots \\
C^k (I-h\Gamma^k)^{m-1} 
\end{bmatrix}\label{eq:observability_matrix_block_zero}
\end{equation}
for all $k \in [m]$.


\begin{thm}\label{thm:Observable_zero}
Under Assumptions~\ref{assume:one}-\ref{assume:four}, if, 
for each $i \in [n]$, $\gamma_i^k$ is a distinct value for all
$k \in[m]$, 
the competing virus model~\eqref{eq:SIR_observation} 
is locally observable at $s_i[t]=0, 
\forall i \in [n]$.
\end{thm}
\textit{Proof:}
From the assumptions $h\sum_{k=1}^m \gamma^k_i\leq 1$, $h>0$, and $\gamma_i^k>0$ for all $i\in [n], k \in [m]$, 
we obtain that 
$1-h\gamma_i^k >0$ 
for all $i\in [n], k \in [m]$. In addition, since we assume that $c_i^k>0$ for all $i\in [n], k \in [m]$, we can conclude that the entries of Eq.~\eqref{eq:observability_matrix_zero}: $c_i^k(1-h\gamma_i^k)\in (0,1)$ for all $i\in [n], k \in [m]$.

We let $\textbf{0}^{n} := \begin{bmatrix}
0 & 0& \cdots & 0
\end{bmatrix}^{1\times n}$ and $\textbf{0}^{0} := \emptyset$.
Consider Eq.~\eqref{eq:observability_matrix_block_zero} and recall that every matrix in it is diagonal; Hence, Eq.~\eqref{eq:observability_matrix_zero} is the concatenation of a set of block diagonal matrices. 
For all $i\in[n]$, the $i$-th row 
of the observability matrix~\eqref{eq:observability_matrix_zero} can be written as: 
\begin{equation}
    \begin{bmatrix}
\textbf{0}^{i-1}& c_i^1 &\textbf{0}^{n-i} &\textbf{0}^{i-1} &c_i^2& \textbf{0}^{n-i}& \cdots & \textbf{0}^{i-1} &c_i^m  &\textbf{0}^{n-i}
\end{bmatrix}\nonumber 
\end{equation}
which is linearly independent with the $(i+ln)$-th row of~\eqref{eq:observability_matrix_zero} for all $l \in [m-1]$:
\begin{equation}
    \begin{bmatrix}
 \textbf{0}^{i-1}& c_i^1(1-h\gamma_i^1)^{l} &\textbf{0}^{n-i}&  
 \cdots & \textbf{0}^{i-1}  & c_i^m(1-h\gamma_i^m)^{l} &\textbf{0}^{n-i}
\end{bmatrix} \nonumber 
\end{equation}
\noindent
under our assumption that, for each $i \in [n]$, $\gamma_i^k$ is a distinct value across all $k\in[m]$. Thus, the observability matrix in Eq.~\eqref{eq:observability_matrix_zero} has full row rank. Since the observability matrix is a square matrix, we conclude that Eq.~\eqref{eq:observability_matrix_zero} is full rank.
Notice that whenever we add another virus to the system model~\eqref{eq:SIR_observation},
we increase the dimension of~\eqref{eq:observability_matrix_zero} from $mn\times mn$ to $(m+1)n \times (m+1)n$ by adding $m$ blocks, 
and the rank of the observability matrix will change from $mn$ to $(m+1)n$, by the same logic as above. 
Therefore, by Lemma \ref{lemma:locally_observable}, the competing virus model in~\eqref{eq:SIR_observation} 
is locally observable at $s_i[t]=0, 
\forall i \in [n]$.
\vspace{-1ex}

\begin{rem}
The assumption in Theorem~\ref{thm:Observable_zero}, namely that, for each $i \in [n]$, $\gamma_i^k$
is a distinct value across every $k \in[m]$, 
can be interpreted as each virus 
having
a different 
recovery rate. 
This assumption is 
reasonable 
as the recovery rate 
represents the inverse of the average duration of an infected individual being sick, and the average amount of time for an individual to recover from different types/strands of viruses varies drastically~\citep{whitley2001herpes}.
\end{rem}

\vspace{-1.5ex}

\section{Simulations}\label{sec:simulation}

\vspace{-2ex}

\begin{table}
\centering
\begin{tabular}{|c | c c c c c|} 
 \hline  
 $\beta_{ij}^c$    & UK & ESP & GER & TUR & RUS \\ [0.5ex] 
 \hline
 UK   & 0.08& 0.15& 0.24& 0& 0.06 \\
 ESP & 0.15& 0.12& 0.13& 0.11& 0  \\
 GER       & 0.24& 0.13& 0.25& 0.05& 0.04  \\
 TUR        & 0& 0.09& 0.05& 0.11& 0.15   \\
 RUS       &  0.06& 0& 0.04& 0.14& 0.09  \\ \hline\hline
 $\gamma_i^c$       &0.15& 0.23& 0.17& 0.25& 0.2  \\
 $x^c[0]$       & 0.02       & 0.04          & 0.03   & 0.01  & 0.03  \\$c^c$       & 0.4       &  0.4          &  0.4   &  0.4  & 0.4   \\ 
 \hline
\end{tabular}
\caption{Network Parameters of SARS-CoV-2 of Figure~\ref{fig:Graph_Europe}.}
\label{table:parameters_COVID}
\end{table}

\begin{table}
\centering
\begin{tabular}{|c | c c c c c|} 
 \hline
 $\beta_{ij}^f$    & UK & ESP & GER & TUR & RUS \\ [0.5ex] 
 \hline
 UK   & 0.02& 0.05& 0.04& 0& 0.01 \\
 ESP & 0.05& 0.06& 0.07& 0.02& 0  \\
 GER       & 0.04& 0.07& 0.04& 0.03& 0.05  \\
 TUR        & 0& 0.03& 0.04& 0.09& 0.07   \\
 RUS       &  0.01& 0& 0.05& 0.07& 0.06 \\ \hline\hline
 $\gamma_i^f$       &0.095& 0.12& 0.1& 0.15& 0.13  \\
 $x^f[0]$       & 0.001       & 0.002          & 0.0035   & 0.002  & 0.001   \\$c^f$       & 0.3       &  0.3          &  0.3   &  0.3  & 0.3\\ 
 \hline
\end{tabular}
\caption{Network Parameters of influenza of Figure~\ref{fig:Graph_Europe}.}
\label{table:parameters_Flu}
\end{table}

\begin{figure}
\centering
\includegraphics[width=.3\textwidth]{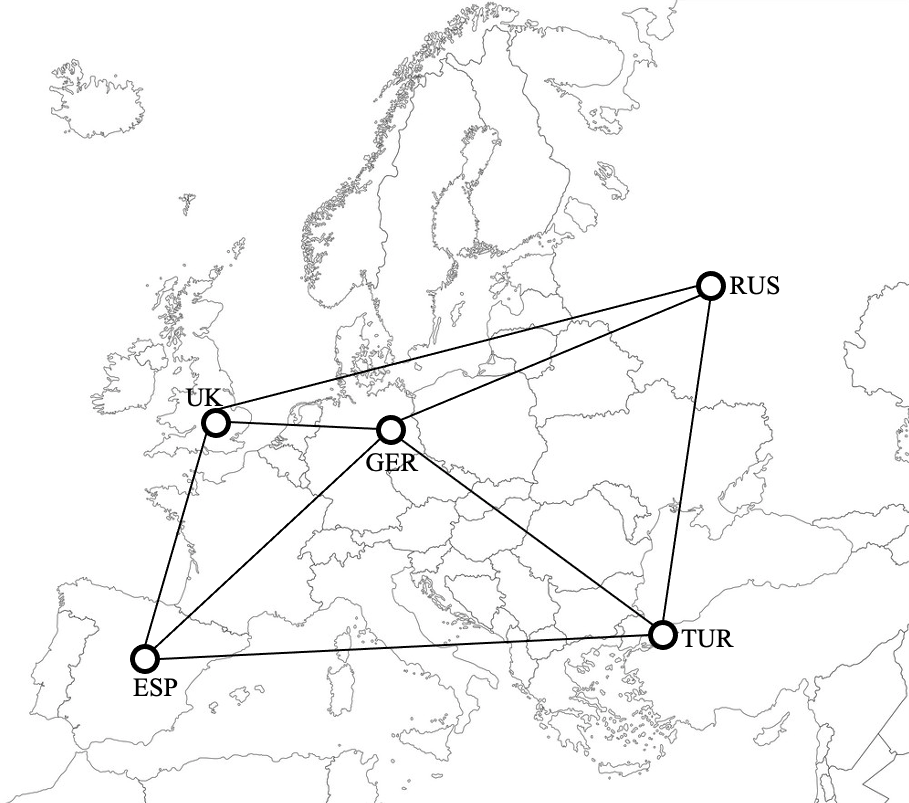}
\caption{Graph topology in the map of Europe analyzed}
\label{fig:Graph_Europe}
\end{figure}

In this section, we consider the special case of two viruses: the SARS-CoV-2 virus and influenza spreading over the network depicted in Figure~\ref{fig:Graph_Europe}. 
In the network, each node represents a major country in Europe: UK, Turkey, Germany, Spain, and Russia, and the edges represent transportation between two node countries. The system parameters are listed in Table~\ref{table:parameters_COVID} and Table~\ref{table:parameters_Flu}.
We choose the SARS-CoV-2 virus and influenza because they cause patients to display similar symptoms such as fever, fatigue, 
headache, etc. It is difficult to distinguish between the two contagions in the early stage of the epidemic without proper testing facilities. 
This section includes no real data; however, the viral spreading parameters are inspired by the behavior of the viruses~\citep{anderson2020will}, namely, the SARS-CoV-2 virus is more contagious than influenza. We also acknowledge that these two viruses are not necessarily competitive; there are cases where people have been infected with both SARS-CoV-2 and influenza \citep{wu2020co}.


\begin{figure}
\centering
\begin{overpic}[width = 0.48\columnwidth]{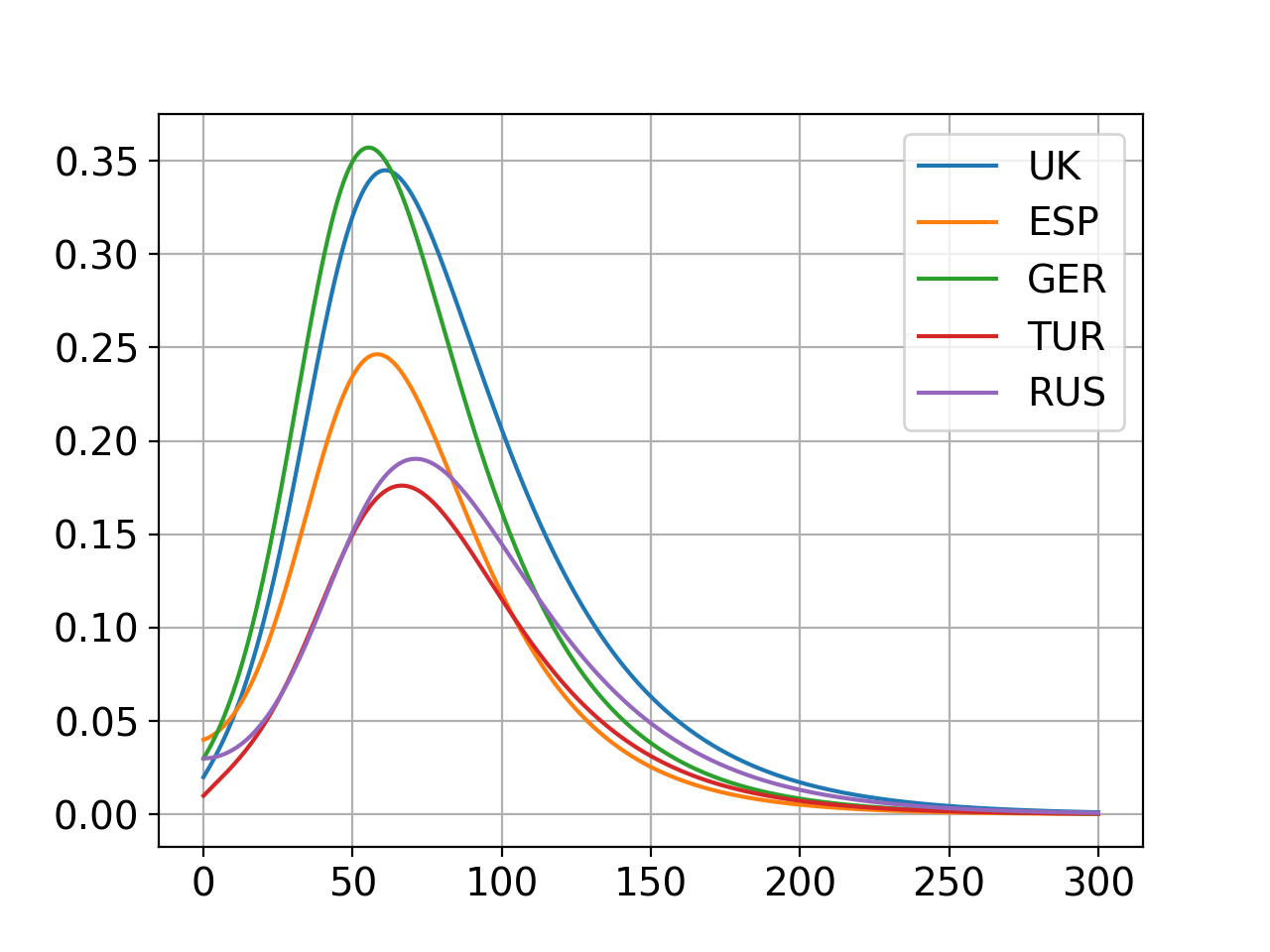}
     \put(-2,35){{\parbox{0.75\linewidth}\footnotesize \rotatebox{90}{\footnotesize$x^c$}
     }}
     \put(50,-1){\footnotesize{\parbox{0.75\linewidth}\footnotesize $t$
     }}\normalsize
   \end{overpic}
\begin{overpic}[width =  0.48\columnwidth]{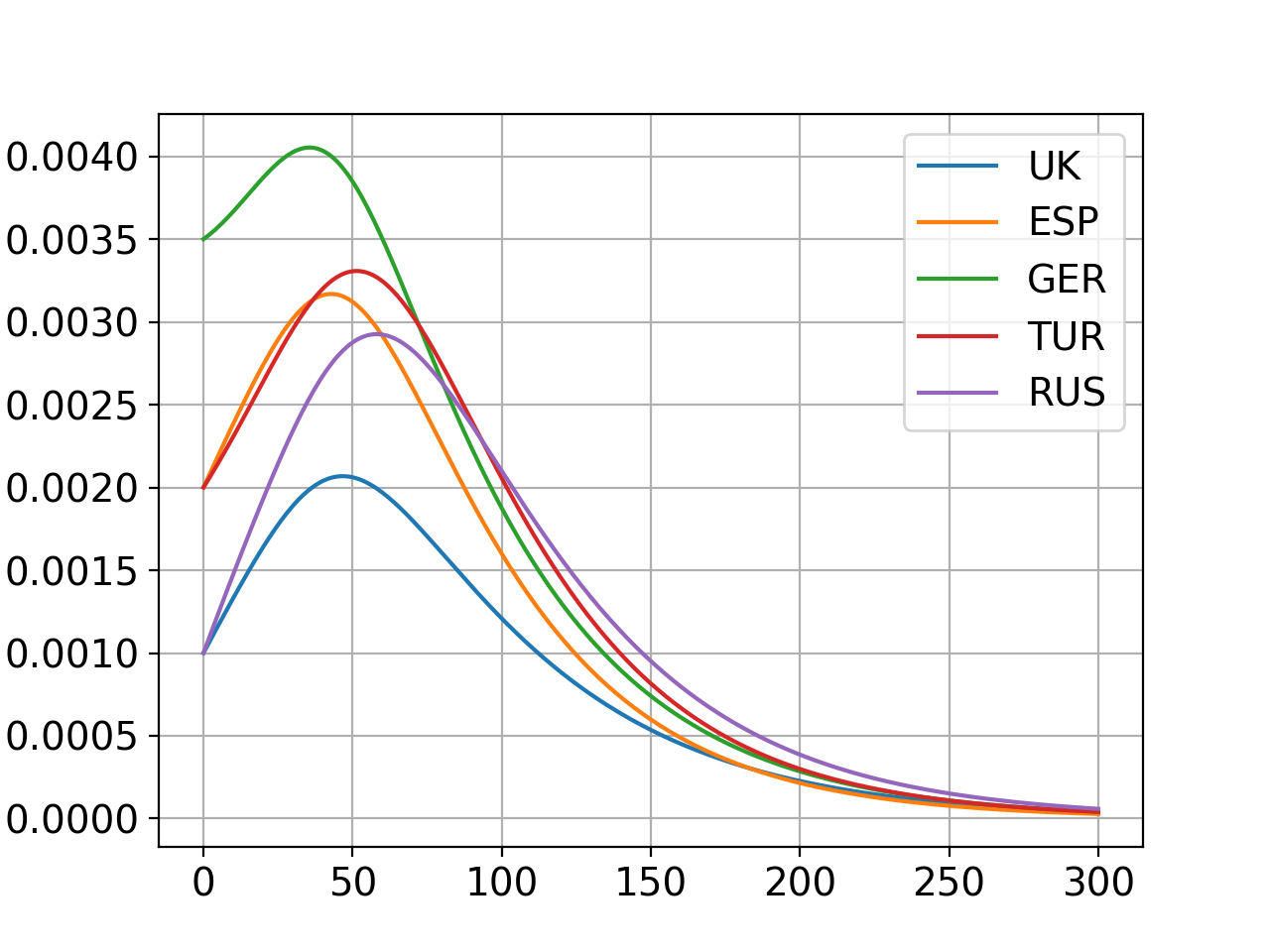}
     \put(-7,35){{\parbox{0.75\linewidth}\footnotesize \rotatebox{90}{\footnotesize$x^f $}
     }}\normalsize
     \put(50,-1){\footnotesize 
    $t$
     }
   \end{overpic}
\caption{Evolution of infection level of SARS-CoV-2 virus in each country (left); Evolution of infection level of influenza in each country (right).}
\label{fig:Evolution}
\end{figure}

\begin{figure}
\centering
\begin{overpic}[width = 0.48\columnwidth]{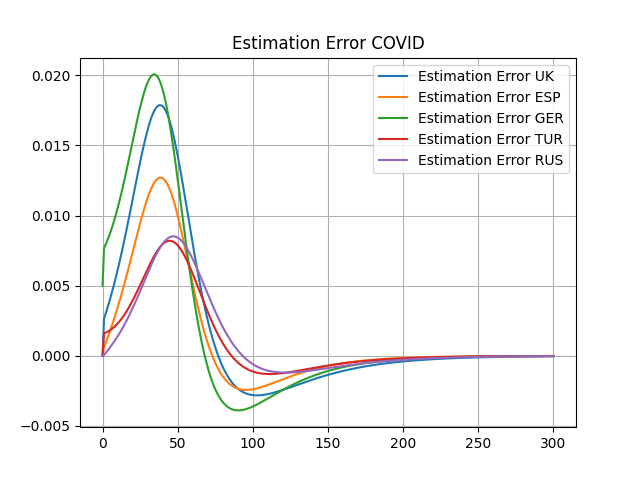}
     \put(-1,25){{\parbox{0.75\linewidth}\footnotesize \rotatebox{90}{\footnotesize$x^c -\hat{x}^c$}
     }}
     \put(50,0){\footnotesize{\parbox{0.75\linewidth}\footnotesize $t$
     }}\normalsize
   \end{overpic}
\begin{overpic}[width =  0.48\columnwidth]{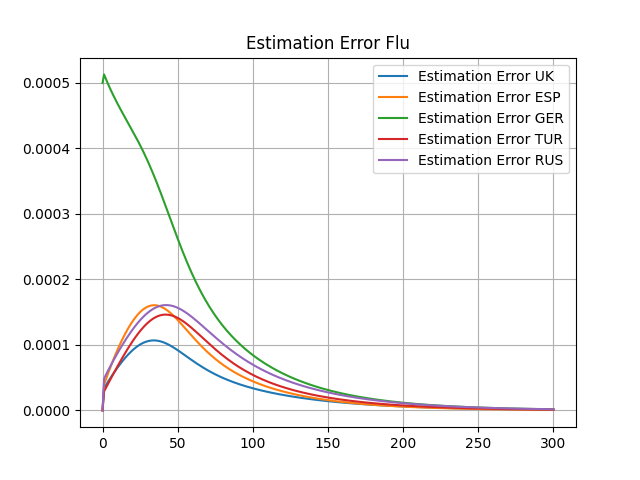}
     \put(-5,25){{\parbox{0.75\linewidth}\footnotesize \rotatebox{90}{\footnotesize$x^f -\hat{x}^f$}
     }}\normalsize
     \put(50,0){\footnotesize 
    $t$
     }
   \end{overpic}
\caption{Estimation error of infection level of SARS-CoV-2 virus in each country (left); Estimation error of infection level of influenza in each country (right).}
\label{fig:Error_nonoise}
\end{figure}

The 
evolution of the infection levels
of both viruses are illustrated in Figure~\ref{fig:Evolution}. 
We estimate the infection level 
by using the following proposed
Luenberger observer:
\vspace{-1ex}
\begin{align}\label{eq:SIR_luenberger_observation_simulation}
    \hat{x}^k_i[t+1] =& \hat{x}^k_i[t] +h\bigg\{\hat{s}_i[t]\sum_{j=1}^n \beta^k_{ij} \hat{x}^k_j[t]-\gamma^k_i \hat{x}^k_i[t]\bigg\} \nonumber  \\&\ \  \ \ 
    +L_i(y_i[t]-\hat{y}_i[t]),
\end{align}
where $\hat{s}_i[t] = 1-\sum_{k=1}^m\hat{x}^k_i[t]-\hat{r}_i[t]$, in which 
the recovered level is estimated through: $\hat{r}_i[t] = h\sum_q^t \sum_{k=1}^m \gamma_i^k \hat{x}_i^k[q]$
at each time step recursively.
We first simulate the state estimation in Figure~\ref{fig:Error_nonoise} and we can see that the estimation error converges to zero before the viruses die out. 
Moreover, the magnitude of the estimation error of each virus in each node is less than $10\%$ of its infection proportion respectively. Hence, the Luenberger observer is an adequate system state estimator for our system model~\eqref{eq:SIR_dynamic}.  

\vspace{-1ex}

\section{Conclusion}\label{sec:conclusion}
\vspace{-2ex}
This paper has investigated the stability and observability of a novel discrete-time networked multi-virus 
SIR
model. 
We have provided a sufficient condition for each virus to converge to zero exponentially. We have then specified a necessary and sufficient condition for the system to be locally observable at $s_i[t]=0, \forall i \in [n]$. In simulation, we utilized a Luenberger state observer to estimate the system states and the results illustrate that the Luenberger observer is suitable for 
state estimation 
of
our new model. 

\vspace{-2ex}
\bibliography{ifacconf}             
                                                   







\end{document}